%% file: skeleton.tex
\documentclass[a4paper,11pt]{article}
\usepackage{pos}
\usepackage{subfig}
\usepackage{wrapfig}
\usepackage{siunitx}
\usepackage{lipsum} 
\usepackage[capitalise]{cleveref}
\usepackage{graphicx}
\usepackage{setspace}
\usepackage{lineno}
\usepackage{anyfontsize}

\title{Update on the Combined Analysis of Muon Measurements from Nine Air Shower Experiments}
 \ShortTitle{Update on the Combined Analysis of Muon Measurements}

\author*[a]{Dennis Soldin}

\affiliation[a]{Bartol Research Institute, Dept. of Physics and Astronomy\\
University of Delaware, Newark, DE 19716, USA}

\forColl{{EAS-MSU}, {IceCube}, {KASCADE-Grande}, {NEVOD-DECOR}, \nobreak{Pierre Auger}, {SUGAR}, {Telescope Array}, and {Yakutsk EAS Array}} % W/O "Collaboration"

\emailAdd{soldin@udel.edu}

\abstract{Over the last two decades, various experiments have measured muon densities in extensive air showers over several orders of magnitude in primary energy. While some experiments observed differences in the muon densities between simulated and experimentally measured air showers, others reported no discrepancies. 

We will present an update of the meta-analysis of muon measurements from nine air shower experiments, covering shower energies between a few PeV and tens of EeV and muon threshold energies from a few 100\,MeV to about 10\,GeV. In order to compare measurements from different experiments, their energy scale was cross-calibrated and the experimental data has been compared using a universal reference scale based on air shower simulations. Above 10\,PeV, we find a muon excess with respect to simulations for all hadronic interaction models, which is increasing with shower energy. For EPOS-LHC and QGSJet-II.04 the significance of the slope of the increase is analyzed in detail under different assumptions of the individual experimental uncertainties.}%8 sigma.}

\FullConference{37$^{\rm{th}}$ International Cosmic Ray Conference (ICRC 2021)\\
		July 12th -- 23rd, 2021\\
		Online -- Berlin, Germany}

\begin{document}
\maketitle

%\begin{linenumbers}

\section{Introduction}

Cosmic rays enter the Earth's atmosphere where they produce \emph{Extensive Air Showers} (EAS) which can be measured at the ground. Although the energy spectrum of cosmic rays has been measured with high precision over many orders of magnitude, the sources of cosmic rays are still unknown, their acceleration mechanism and mass composition are uncertain, and several features observed in the energy spectrum are not well understood~\cite{Kampert:2012mx,BeckerTjus:2020xzg}. 

The main challenge lies in the measurements of the muon content in air showers, which have been performed by many experiments over the last 20 years. Various experiments reported discrepancies in the number of muons in simulated and observed air showers, such as the \mbox{HiRes-MIA}~\cite{AbuZayyad:1999xa} and NEVOD-DECOR~\cite{Bogdanov:2010zz,Bogdanov:2018sfw} collaborations, as well as the Pierre Auger Observatory~\cite{Aab:2014pza,Aab:2016hkv} (Auger), Telescope Array~\cite{Abbasi:2018fkz} (TA), and SUGAR~\cite{Bellido:2018toz}. In contrast, no discrepancies in the average muon densities were observed by EAS-MSU~\cite{Fomin:2016kul}, the Yakutsk EAS array~\cite{Glushkov}, and KASCADE-Grande~\cite{Apel:2017thr}. KASCADE-Grande, however, reported differences in the muon number evolution with the zenith angle with respect to model predictions.

\begin{figure}[b]
  \vspace{-2em}
 
  \mbox{\hspace{-.9em}\subfloat[]{\includegraphics[width=.375\textwidth]{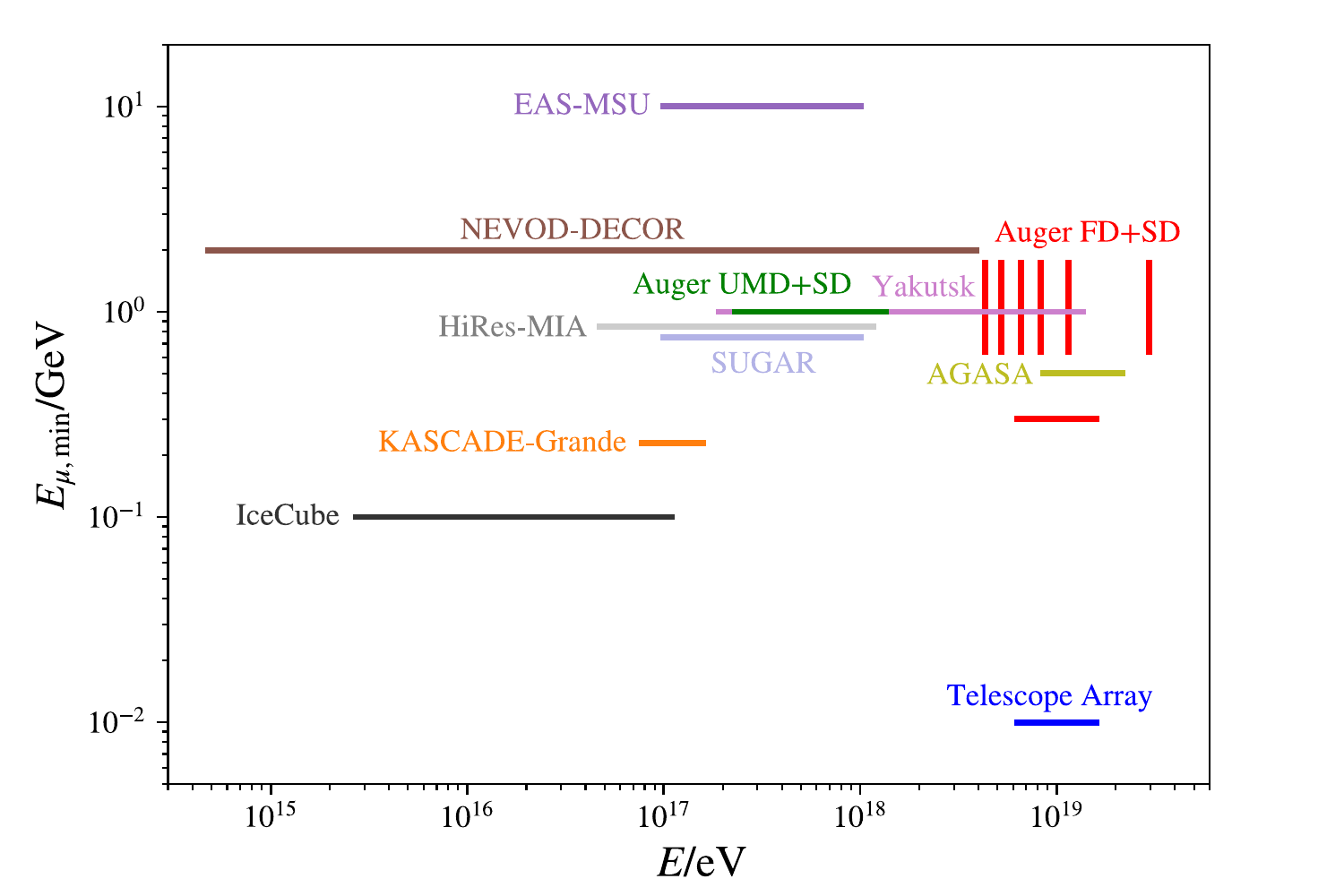}}\hspace{-1.5em}
  \subfloat[]{\includegraphics[width=.375\textwidth]{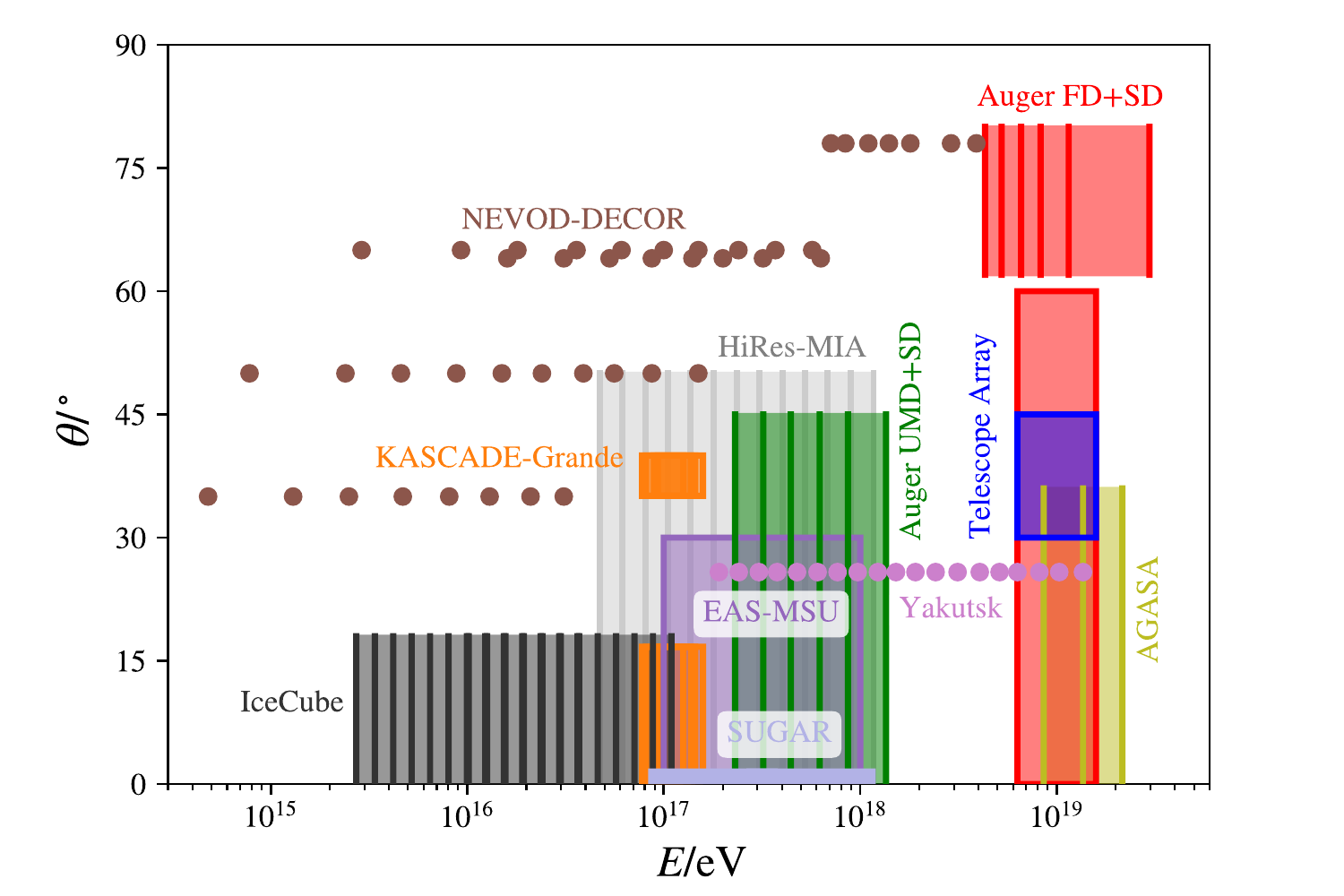}}\hspace{-1.5em}
  \subfloat[]{\includegraphics[width=.375\textwidth]{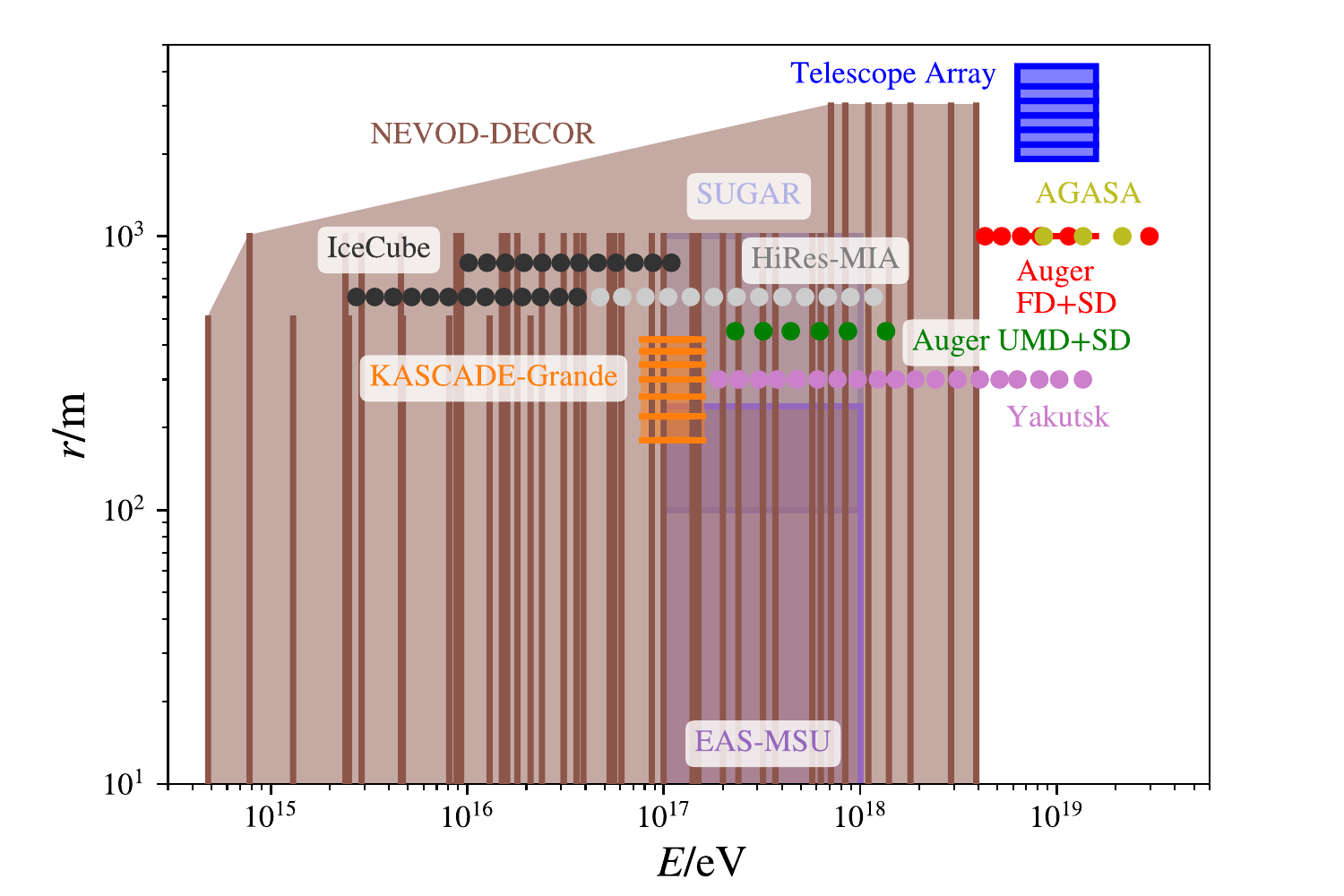}}
  }
  \vspace{-0.5em}
  
  \caption{Phase space of EAS experiments which have reported measurements of the muon density. Points
and lines indicate a measurement in a narrow bin of the parameter, while boxes indicate integration over a parameter range. Figure (a) shows the muon energy thresholds, $E_{\mu,\mathrm{min}}$, (b) the zenith angle range, $\theta$, and (c) the lateral distances, $r$, of the muon density measurements, as a function of the EAS energy, $E$.}
  \label{fig:phase_space}%
  \vspace{-1.em}
  
\end{figure}

In this article, we present an update of the meta-analysis of global measurements of the lateral muon density by multiple EAS experiments which was previously reported in Refs.~\cite{Dembinski:2019uta,Cazon:2020zhx}. This update includes new data from Auger~\cite{Aab:2021zfr} and its Underground Muon Detectors~\cite{Aab:2020frk} (UMD), %, from KASCADE-Grande~\cite{Arteaga-Velzquez:2020zza}, 
and from the IceCube Neutrino Observatory~\cite{Soldin:2021} (IceCube). In addition, data from the AGASA experiment~\cite{Gesualdi:ICRC2021} is included for the first time and systematic studies of the energy-dependent trend of the muon discrepancies are discussed in detail. An overview of further measurements of the muon content in EAS is beyond the scope of this work and can be found in Refs.~\cite{Dembinski:2019uta,Albrecht:2021yla}.

\section{Measurements of the Muon Lateral Density}

The lateral density of muons measured at the ground depends on various parameters: cosmic-ray energy, $E$, zenith angle, $\theta$, shower age (vertical depth, $X$, and zenith angle), lateral distance, $r$, from the shower axis, and energy threshold, $E_{\mu,\mathrm{min}}$, of the muon detectors. The parameter space covered by the experiments considered in this meta-analysis is shown in~\cref{fig:phase_space}. Due to different experimental conditions and analysis techniques, a direct comparison of the muon measurements is not possible. Instead, to compare different results in a meaningful way, the measurements of each experiment have to be compared to EAS simulations in the same observation conditions in terms of a data/MC ratio. Therefore, the measurements of the muon density from different experiments are converted to the \emph{z-scale},
\begin{equation}
\label{eq:z-values}
z=\frac{\ln \langle N_\mu^\mathrm{det}\rangle-\ln\langle N_{\mu,\mathrm{p}}^\mathrm{det}\rangle}{\ln\langle N_{\mu,\mathrm{Fe}}^\mathrm{det}\rangle-\ln\langle N_{\mu,\mathrm{p}}^\mathrm{det}\rangle},
\end{equation}
where $\langle N_\mu^\mathrm{det}\rangle$ is the average muon density estimate as seen in the detector, while $\langle N_{\mu,\mathrm{p}}^\mathrm{det}\rangle$ and $\langle N_{\mu,\mathrm{Fe}}^\mathrm{det}\rangle$ are the simulated average muon densities for proton and iron showers after a full detector simulation. The z-scale is constructed such that an observation of the muon density of $z=0$ is consistent with a simulated proton shower and $z=1$ for a simulated iron shower. %
%The z-values range from $z=0$ for pure proton showers to $z=1$ for pure iron showers if there is no discrepancy between experimental data and simulations. 
A dedicated discussion on the properties and calculation of the z-scale and its uncertainties can be found in Ref.~\cite{Gesualdi:ICRC2021}.%, and $\langle \ln(A)\rangle = z\cdot\ln(56)$ is the average logarithmic mass.

\begin{figure}[tb]
  \vspace{-2.em}
  
  \mbox{\hspace{-2.3em}\includegraphics[width=1.1\textwidth]{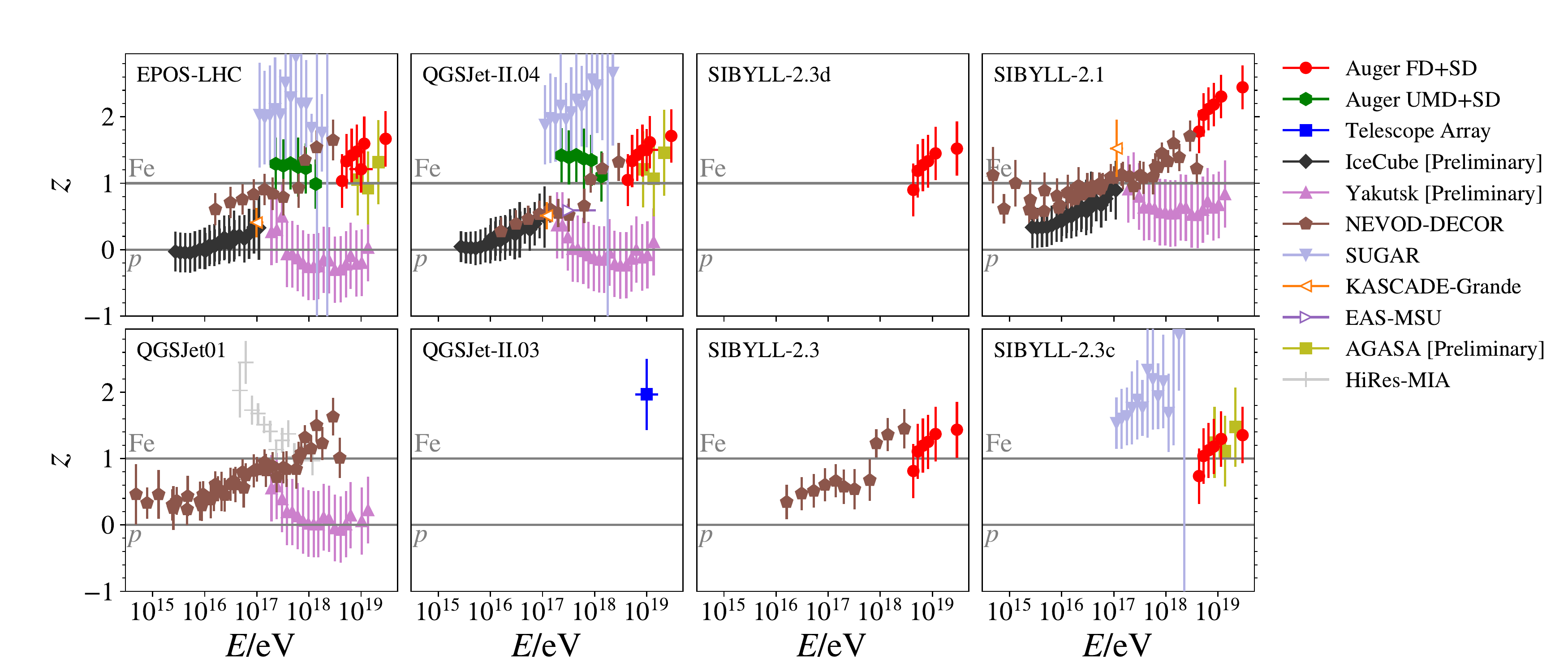}}
  \vspace{-1.5em}
  
  \caption{Muon density measurements converted to the z-scale, as defined in \cref{eq:z-values}, for different hadronic interaction models. When corresponding simulations are missing for an experiment, no points can be shown. Error bars show statistical and systematic uncertainties added in quadrature.}
  \label{fig:rho_mu_raw}%
  \vspace{-1.em}
  
\end{figure}

The z-values obtained by nine air shower experiments~\cite{Bogdanov:2010zz,Bogdanov:2018sfw,Bellido:2018toz,Fomin:2016kul,Glushkov,Aab:2014pza,Apel:2017thr,Aab:2021zfr,Aab:2016hkv,Abbasi:2018fkz,Aab:2020frk,Soldin:2021,Gesualdi:ICRC2021} %Aab:2014pza 
are shown in \cref{fig:rho_mu_raw}. Depending on the experiment, the distributions are given for the \emph{post-LHC} hadronic interaction models EPOS-LHC~\cite{Pierog:2013ria}, QGSJet-II.04~\cite{Ostapchenko:2013pia}, and Sibyll~2.3(c/d)~\cite{Riehn:2017mfm,Engel:2019dsg}, and for the \emph{pre-LHC} models %QGSJet-II.02~\cite{Ostapchenko:2005nj}, 
QGSJet01 and QGSJet-II.03~\cite{Ostapchenko:1993}, and Sibyll~2.1~\cite{Ahn:2009wx}. For all models the data lies between the expectations for proton and iron showers up to energies of about $10^{17}\,\mathrm{eV}$. However, at higher energies all data except Yakutsk suggest an unphysical mass composition heavier than iron.

\subsection{Energy Scale Offsets and Cross-Calibration}

According to the \emph{Matthews-Heitler model}~\cite{Matthews:2005sd} the number of muons in EAS, $N_\mu$, depends on the energy, $E$, and mass, $A$, of the initial cosmic ray as% $N_\mu=A^{1-\beta}\cdot (E/C)^\beta$,
\begin{equation}
\label{eq:heitler_model}
    %N_\mu=A^{1-\beta}\cdot \left(\frac{E}{C} \right)^\beta,
    N_\mu=A^{1-\beta}\cdot (E/\upxi_C)^\beta ,
\end{equation}
with power-law index $\beta\simeq 0.9$ and energy constant, $\upxi_C$. This causes two experiments with an energy-scale offset of $20\%$, for example, to have an $18\%$ offset in the data/MC ratios because measurements are compared to EAS simulated at different apparent energies. To compare muon measurements between experiments, energy-scale offsets therefore need to be taken into account.

\vspace{-1em}
%\end{linenumbers}

%
\newpage

\begin{wrapfigure}{r}{0.35\textwidth}
%\vspace{-2.5em}

\mbox{\hspace{-0em}\includegraphics[width=0.34\textwidth]{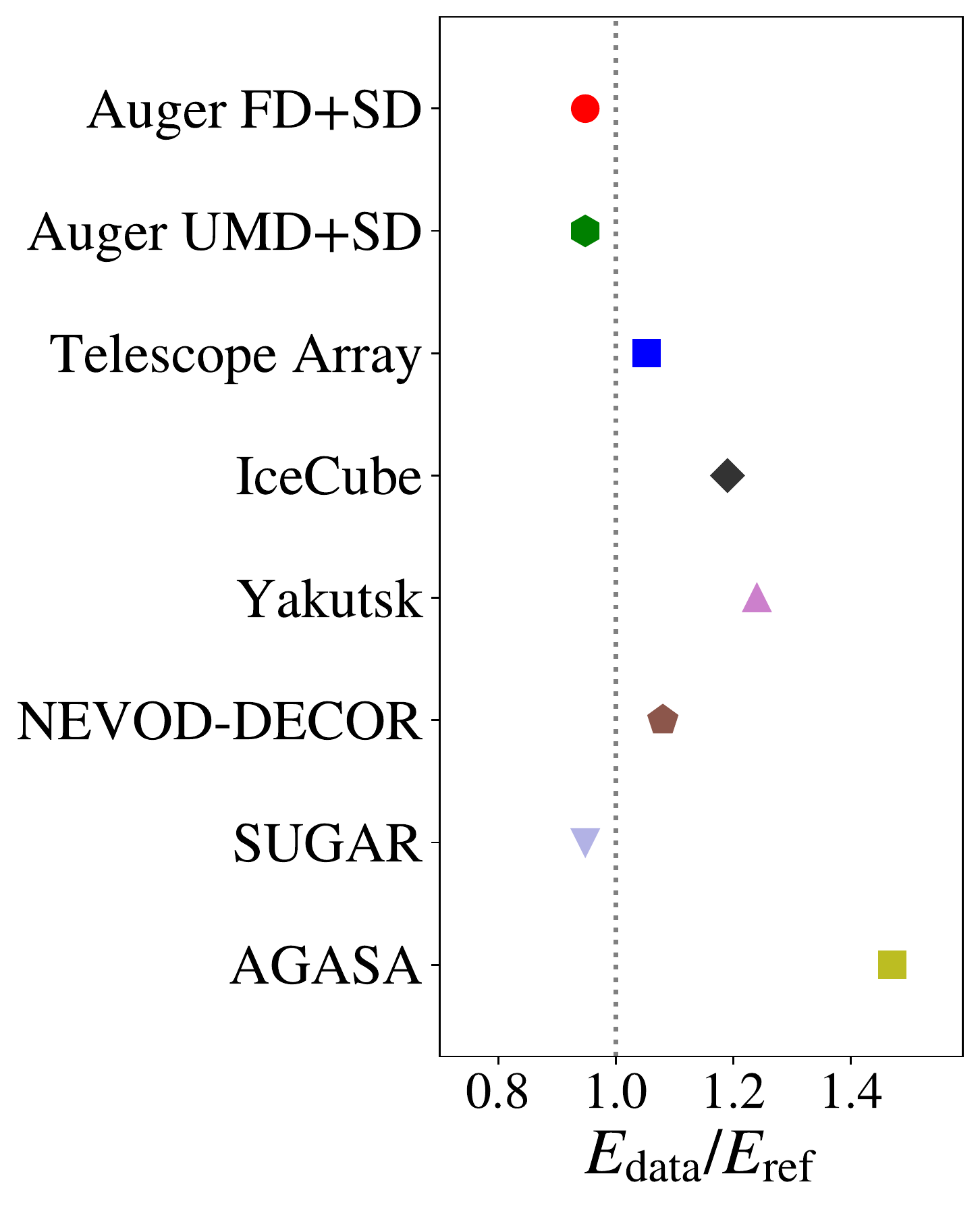}}
 % \vspace{-.5em}
  
\caption{Energy-scale adjustment factors obtained from the
cross-calibration described in Ref.~\cite{Dembinski:2015xtn}.}
\label{fig:scaling_factors}
\vspace{-0.5em}

\end{wrapfigure}

%\vspace{-1em}

Assuming that the cosmic ray flux is a universal reference and that all deviations in measured fluxes between different experiments arise from energy scale offsets, a relative scale $E_\mathrm{data}/E_\mathrm{ref}$ can be determined for each experiment such that the all-particle fluxes match~\cite{Dembinski:2015xtn}. The relative energy-scale shift between Auger and TA has been found to be $10.4\%$~\cite{Deligny:2020gzq} and the reference energy scale, $E_\mathrm{ref}$, used in this work is placed between the two experiments, as shown in~\cref{fig:scaling_factors}. The scaling factors for the other experiments are obtained from the Global Spline Fit (GSF) flux model~\cite{Dembinski:2015xtn} which also uses cross-calibration internally, with a reference energy scale $E_\mathrm{ref,GSF}/E_\mathrm{ref} = 0.948/0.880 \simeq 1.08$. Using the adjustment factors from \cref{fig:scaling_factors}, the z-values are energy cross-calibrated as described in detail in Ref.~\cite{Dembinski:2019uta} and the individual uncertainties of each experiment are adjusted by removing the contribution from the energy-scale. However, the reference energy-scale, after cross-calibration, has a remaining uncertainty of at least $10\%$, causing potential shifts of the z-values by about $\pm 0.25$. No cross-calibration factor can be given for KASCADE-Grande because the cosmic ray flux is computed using a different energy estimator to which this method can not be applied~\cite{Dembinski:2019uta}. For EAS-MSU, no all-particle flux is available for cross-calibration. %SUGAR uses the flux from Auger in its computation of the data/MC ratio and therefore has the same energy-scale% adjustment factor. 

%\begin{linenumbers}

\begin{figure}[b]
  \vspace{-2.5em}
  \mbox{\hspace{-2.3em}\includegraphics[width=1.1\textwidth]{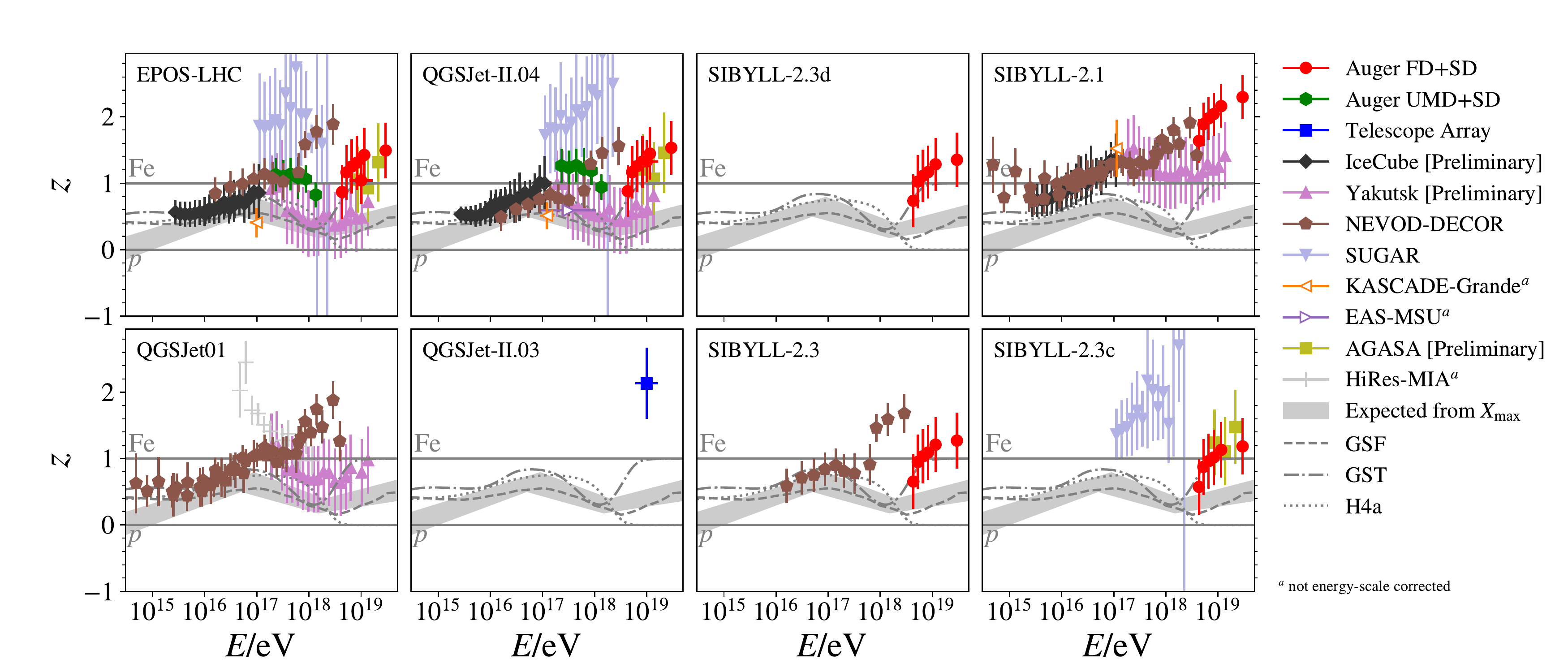}}
  \vspace{-1.5em}
  
  \caption{Combined data from \cref{fig:rho_mu_raw} after applying energy-scale cross-calibration, as described in the text. The data of KASCADE-Grande and EAS-MSU cannot be cross-calibrated and are only included for comparison. Shown for comparison are z-values expected for a mixed composition from optical measurements ($X_\mathrm{max}$), based on an update of Ref.~\cite{Kampert:2012mx}, and from the flux models GSF~\cite{Dembinski:2015xtn}, GST~\cite{Gaisser:2013bla}, and H4a~\cite{Gaisser:2011cc}.}
  \label{fig:rho_mu_rescaled}%
  \vspace{-1em}
  
\end{figure}

The resulting z-values, after applying the energy-scale cross-calibration, are shown in \cref{fig:rho_mu_rescaled}, where a remarkably consistent picture is obtained. The measurements are in agreement with simulations based on the post-LHC hadronic interaction models, EPOS-LHC and QGSJetII.04, up to about a few $10^{16}\,\mathrm{eV}$, within the expectation from measurements of the maximum shower depth, $X_\mathrm{max}$, and uncertainties. However, at higher energies, an increasing muon excess with respect to simulations is observed for all models, suggesting a mass composition heavier than iron. %The slope of this increase in z per decade in energy is 0.22 to 0.35 for EPOS-LHC and QGSJet-II.04.%, with 8 σ significance.

\iffalse
\begin{equation}
    z_\mathrm{ref}=z_\mathrm{data}+\frac{\beta\cdot\ln(E_\mathrm{data}/E_\mathrm{ref})}{\ln(\rho_{\mu,\mathrm{Fe}})-\ln(\rho_{\mu,\mathrm{p}})}
\end{equation}
with $\beta=1-\left(\ln(\rho_{\mu,\mathrm{Fe}})-\ln(\rho_{\mu,\mathrm{p}})\right)/\ln(56)$

\fi

\subsection{Energy-Dependent Trend}

\begin{figure}[tb]
  \centering
  \vspace{-1.5em}
  
  \mbox{\hspace{-0.5em}%
    \includegraphics[width=0.45\textwidth]{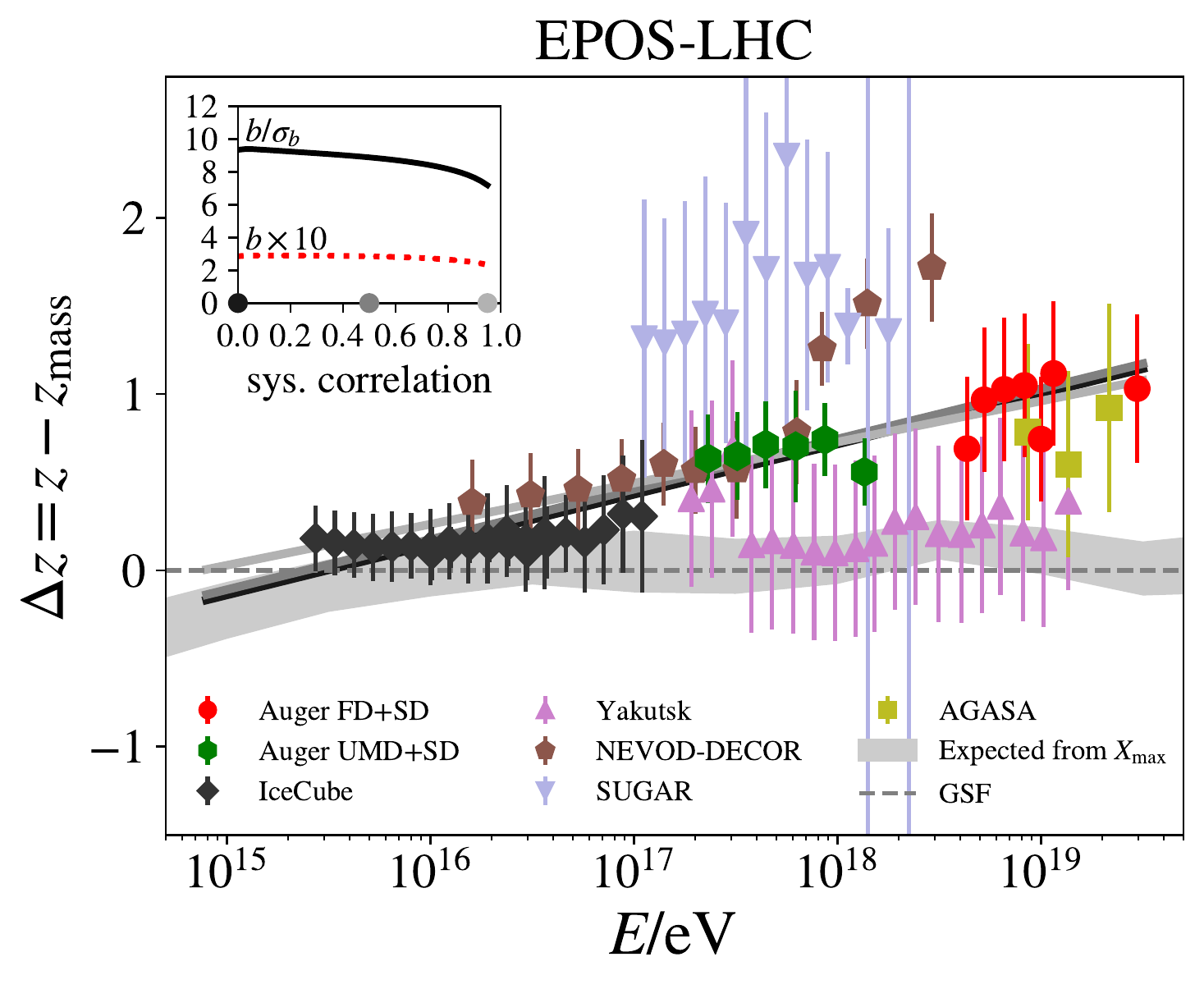}\quad
    \includegraphics[width=0.45\textwidth]{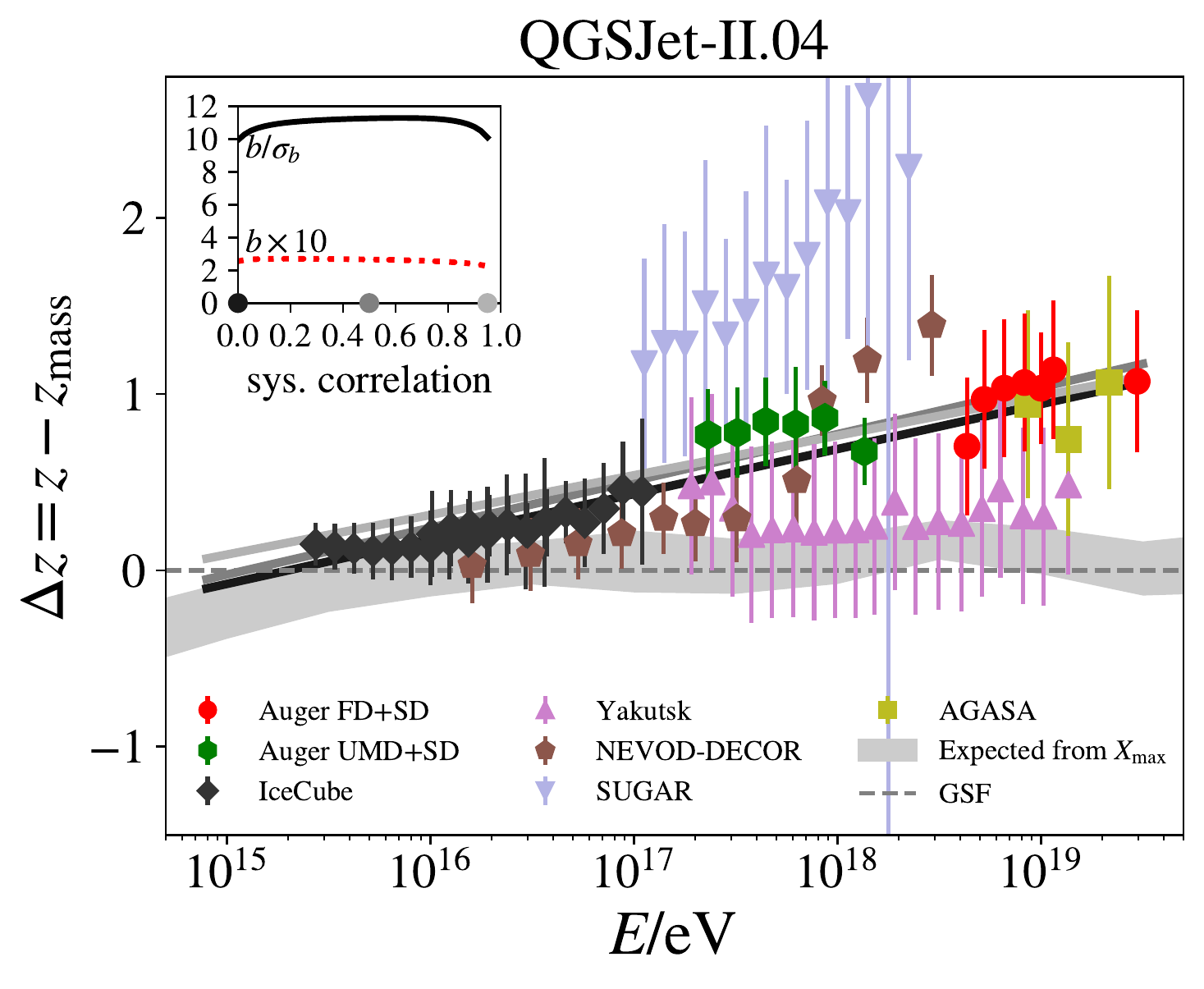}%
    %\label{fig:qgsjet_fit}%
  }
  \vspace{-.5em}
  
  \caption{Linear fits to the $\Delta z=z-z_\mathrm{mass}$ distributions, as described in~\cref{eq:fit}. Shown in the inset are the slope, $b$, and its deviation from zero in standard deviations for an assumed correlation of the point-wise uncertainties within each experiment. Examples of the fits are shown for a correlation of $0.0$, $0.5$, and $0.95$.}% in varying shades of gray.}
  \label{fig:main_fits}%
  \vspace{-1em}
  
\end{figure}

 In order to systematically quantify the energy-dependent trend observed in the muon measurements shown in \cref{fig:rho_mu_rescaled}, the mass composition dependence expected from \cref{eq:heitler_model} needs to be taken into account. If the measured z-values follow $z_\mathrm{mass}$ as expectated from $X_\mathrm{max}$ measurements, the model describes the muon density at the ground consistently. Subtracting $z_\mathrm{mass}$ is thus expected to remove the effect of the changing mass composition. The resulting $\Delta z=z-z_\mathrm{mass}$ distributions are shown in \cref{fig:main_fits} for EPOS-LHC and QGSJet-II.04 with $z_\mathrm{mass}$ determined from the GSF flux model~\cite{Dembinski:2015xtn}. A simple linear function of the form
\begin{equation}
 \label{eq:fit}
 \Delta z_\mathrm{fit} = a+b\cdot\log_{10}(E/10^{16}\mathrm{eV})
\end{equation}

\begin{figure}[b]
  \centering
  \vspace{-1.5em}
  
  \mbox{\hspace{-0.5em}
    \includegraphics[width=0.45\textwidth]{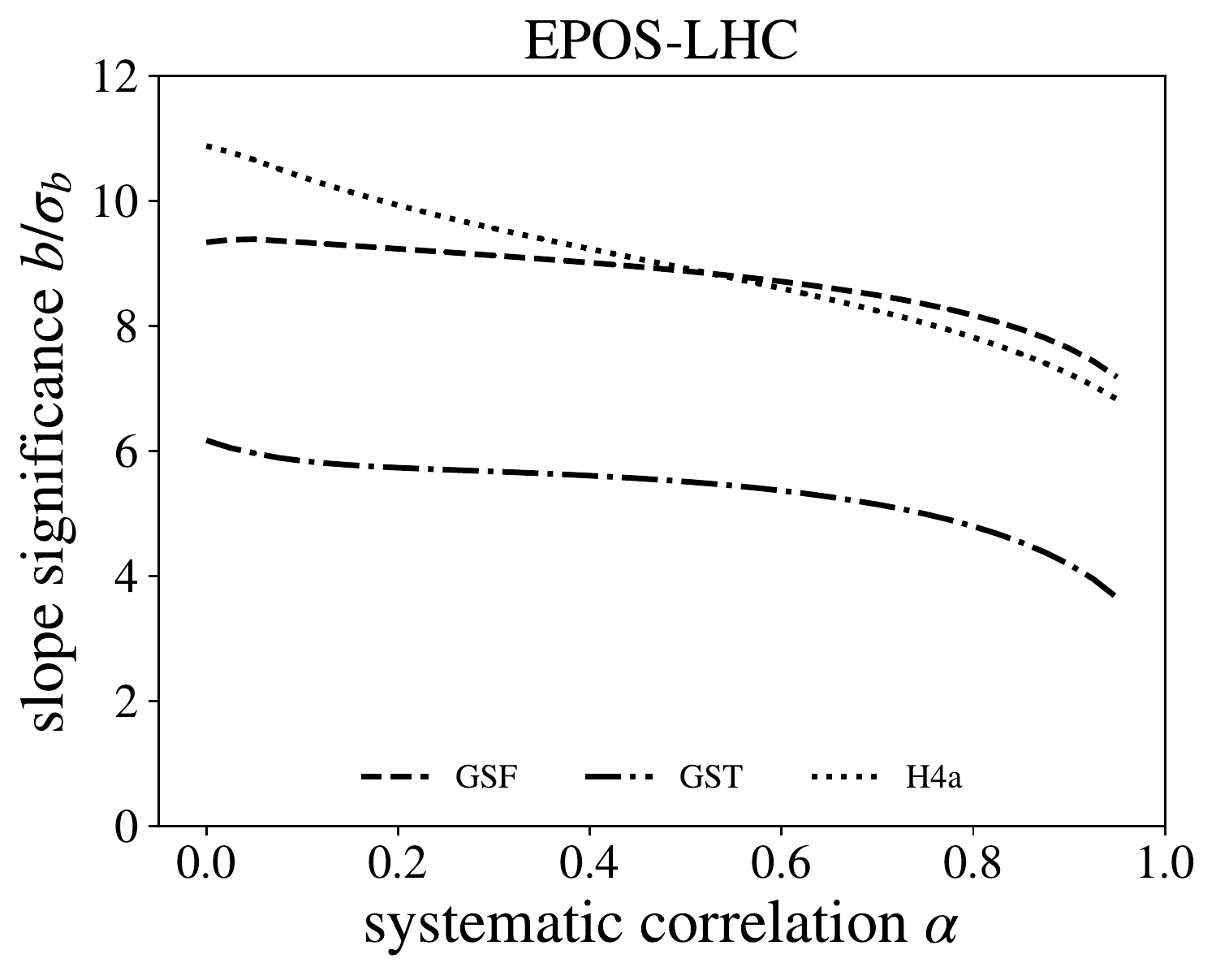}\quad
    \includegraphics[width=0.45\textwidth]{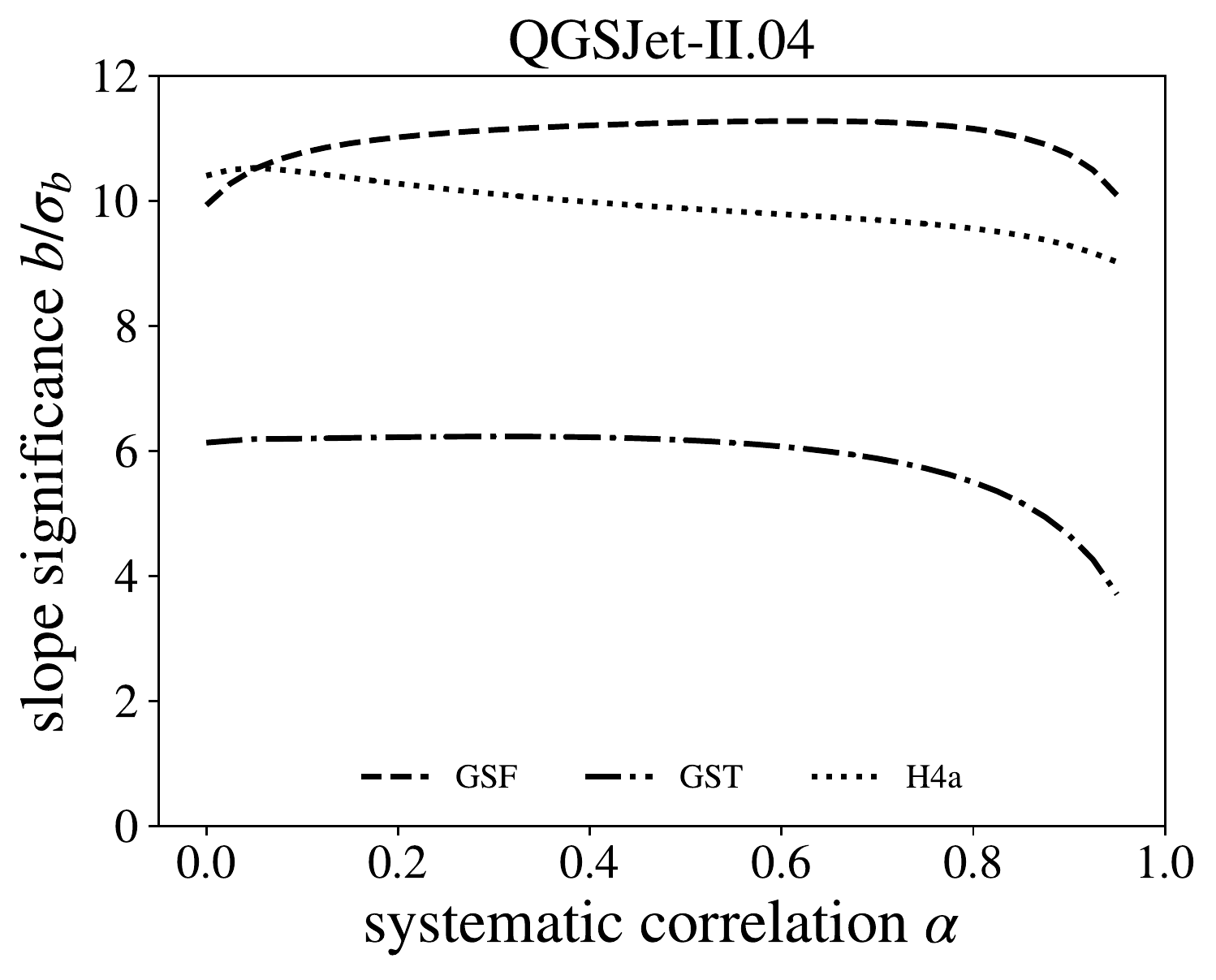}%
  }
  \vspace{-.5em}
  
  \caption{Significance of the deviation of the slope, $b$, in \cref{eq:fit} from zero as a function of an assumed correlation, $\alpha$, of the uncertainties within each experiment. The significances are shown for $\Delta z$ with $z_\mathrm{mass}$ determined from the GSF, GST, and H4a models.}%, and based on the expectation from $X_\mathrm{max}$.}
  \label{fig:mass_dependence}%
  \vspace{-1em}
  
\end{figure}

\noindent is fit to the data, where $a$ and $b$ are free parameters. To account for (the unknown) correlated uncertainties in the experimental data, the least-squares method described in Ref.~\cite{Dembinski:2019uta} is used. It assumes a correlation factor, $\alpha$, between data points belonging to the same data set. The fit is repeated for values of $\alpha$ between $0$ and $0.95$. To adjust for over/under-estimated uncertainties, the raw result, $\sigma_b^\mathrm{raw}$, is re-scaled with the $\chi^2$ value and the degrees of freedom, $n_\mathrm{dof}$, of the fit as $\sigma_b=\sigma_b^\mathrm{raw}\cdot \sqrt{\chi^2/n_\mathrm{dof}}$, as described in Ref.~\cite{Dembinski:2019uta}. For EPOS-LHC, the resulting slope ranges from $b=0.23\pm 0.03$ up to $b=0.29\pm 0.03$ for $\alpha=0.95$ and $\alpha=0.0$, respectively. For QGSJet-II.04, it ranges from $b=0.22\pm 0.02$ up to $b=0.25\pm 0.03$ for $\alpha=0.95$  and $\alpha=0.0$. The significances of the deviations from $b=0.0$ are around $8\sigma$ for EPOS-LHC and above $10\sigma$ for QGSJet-II.04.% for any assumed correlation.

To study the influence of the underlying mass assumption, the fits in $\Delta z=z-z_\mathrm{mass}$ are repeated with $z_\mathrm{mass}$ obtained from the GST~\cite{Gaisser:2013bla} and H4a~\cite{Gaisser:2011cc} flux models which are also based on fits to experimental data. %This is also done assuming that $z_\mathrm{mass}$ follows the average of the expectation from optical $X_\mathrm{max}$ measurements. 
The resulting significances for correlations between $0.0$ and $0.95$ are shown in \cref{fig:mass_dependence}. While the significances based on the H4a model increase, the GST model always yields smaller significances. However, as shown in \cref{fig:rho_mu_rescaled}, the GST model predicts a heavier cosmic ray mass composition which is in strong tension with measurements of $X_\mathrm{max}$ over the vast majority of the energy range considered here. %The significances based on the average of the expectation from $X_\mathrm{max}$ are approximately $6\sigma$, however, the exact functional form of the distribution within the error band depicted in \cref{fig:main_fits} is unknown. Thus, the GSF model, which agrees with the expectation from $X_\mathrm{max}$ over most of the energy range, is used as the reference mass composition for further analysis.
This confirms the choice of the GSF model which shows the best overall agreement with optical measurements of the mass composition. In addition, the effect of the choice of the reference energy scale, $E_\mathrm{ref}$ in \cref{fig:scaling_factors}, was also studied and found to be negligible. %, as shown in \cref{fig:ref_scale_sigma}. 
%For any choice of reference energy scale in the range of the experiments considered in this analysis the significances of the fits remain to be around $7\sigma$ for EPOS-LHC and $9\sigma$ for QGSJet-II.04.

\subsection{N-1 Tests}

\begin{figure}[tb]
  \vspace{-1.5em}
  
  \mbox{\hspace{0.8em}\includegraphics[width=.97\textwidth]{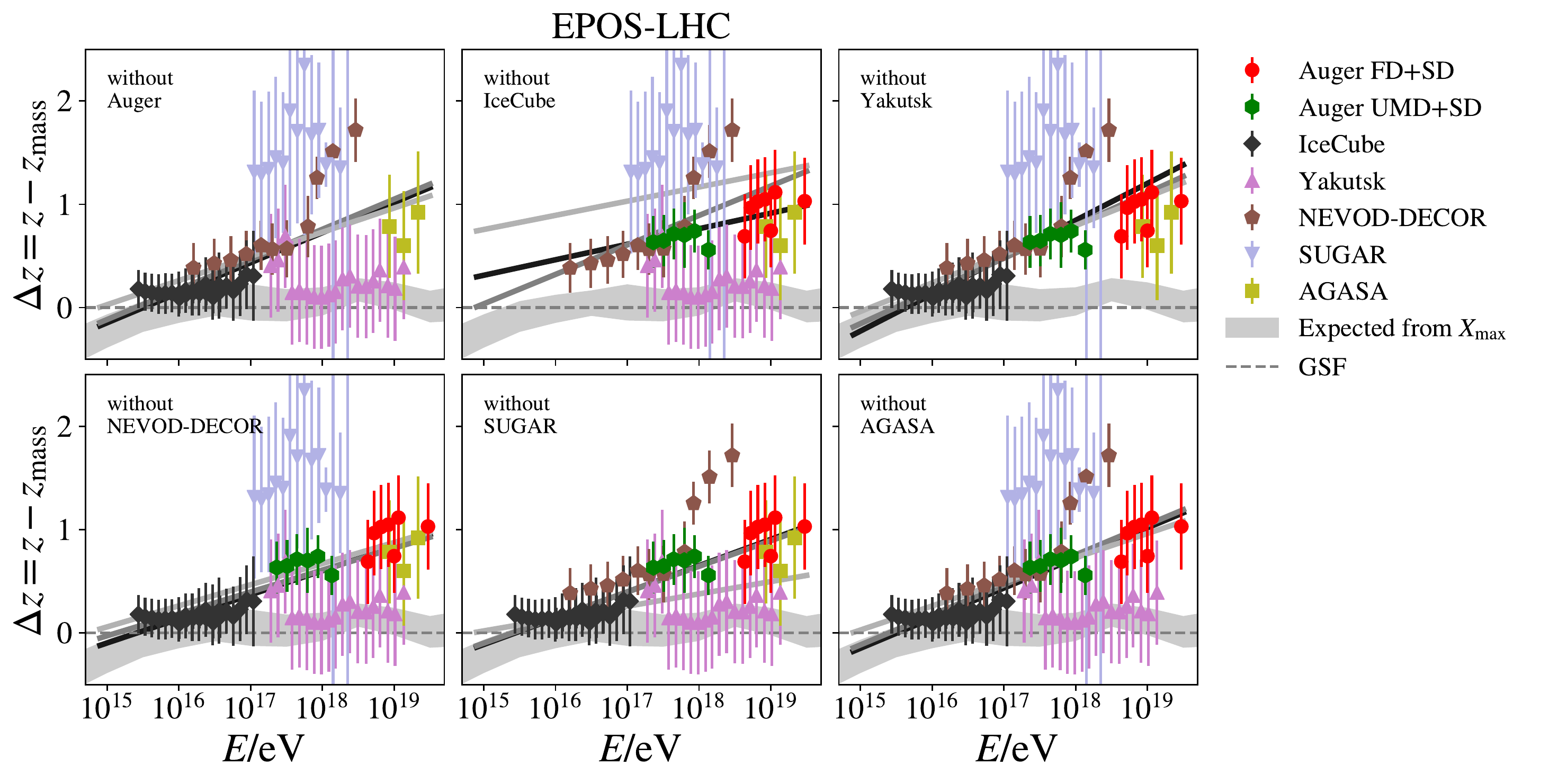}}
  \vspace{-.8em}
  
  \caption{Linear fits to the data points as described in~\cref{eq:fit} with individual experiments excluded from the data. Examples of the fits are shown for systematic correlations of $0.0$, $0.5$, and $0.95$.}
  \label{fig:N-1}%
  \vspace{-1em}
  
\end{figure}

\begin{figure}[tb]
  \centering
  \vspace{-1.2em}
  
  \mbox{-\hspace{0.5em}
    \includegraphics[width=0.45\textwidth]{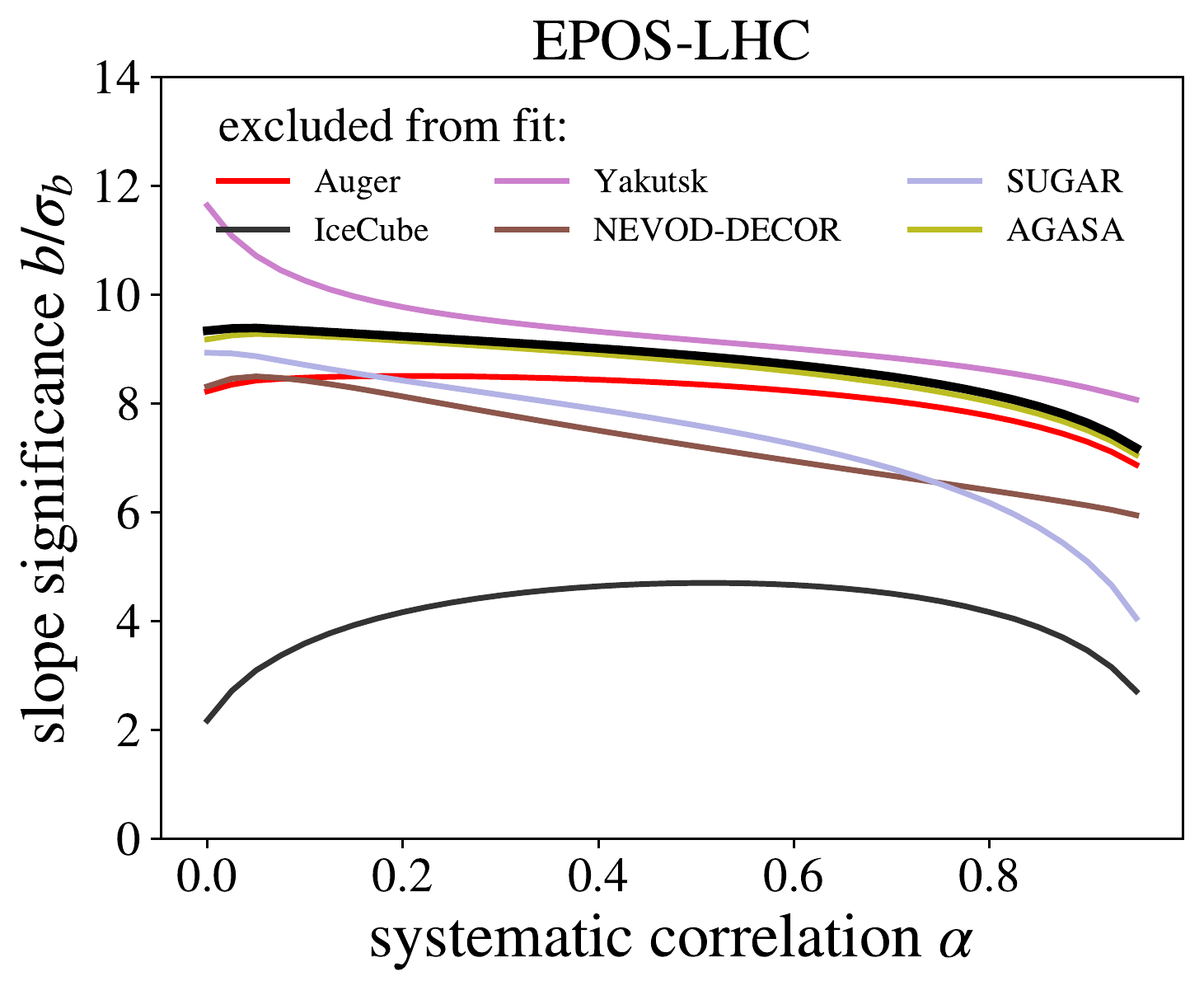}\quad
    \includegraphics[width=0.45\textwidth]{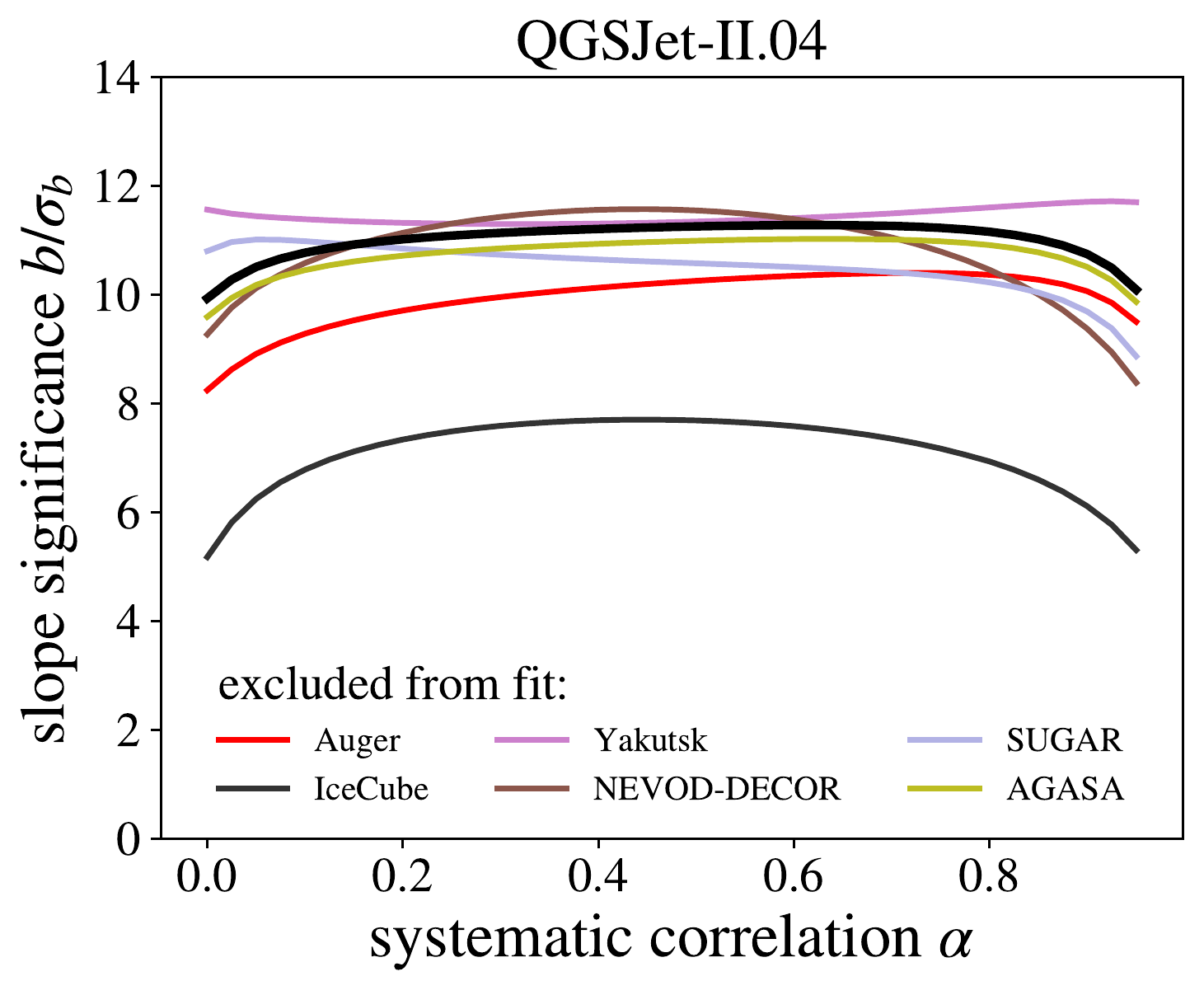}%
  }
  %\vspace{-.5em}
  
  \caption{Significance of the deviation of the slope, $b$, in \cref{eq:fit} from zero as a function of an assumed systematic correlation of the uncertainties within each experiment, $\alpha$. Black lines show the result from \cref{fig:main_fits} and colored lines the results where individual experiments are excluded from the fit (see text for details).}
  \label{fig:N-1_sigma}%
  \vspace{-1.em}
  
\end{figure}

To study the contribution from each individual experiment to the significances of the fit slopes to the combined data, systematic $N-1$ tests are performed where data from one experiment at a time is excluded from the fit. The resulting fits for EPOS-LHC are shown in~\cref{fig:N-1} and the corresponding significances for the models EPOS-LHC and QGSJet-II.04 are depicted in~\cref{fig:N-1_sigma} for systematic correlations $\alpha$ between $0.0$ and $0.95$. The significances of the fit slopes remain above $5\sigma$ when excluding most experiments. However, lower significances can be observed for EPOS-LHC, for extreme correlations, when removing data from IceCube, and to some extent when removing data from SUGAR. In contrast, excluding the measurements by Yakutsk causes an increase of the significances, in particular towards very small correlations. This indicates that measurements from Yakutsk, which are in strong tension with other data in the same energy region, have a stronger influence on the fit result if data from IceCube is removed. %  or NEVOD-DECOR are
%This indicates that by removing data from IceCube or NEVOD-DECOR, measurements from Yakutsk, which are in strong tension with other muon measurements in the same energy region, become more important in the total significance. 
These effects are more pronounced for EPOS-LHC compared to QGSJet-II.04. This is due to a smaller scatter of the data for QGSJet-II.04, which causes smaller uncertainties in the slope, $b$, of the fit through the $\chi^2/n_\mathrm{dof}$ re-scaling.

\vspace{-.5em}

\section{Conclusions}

An update of the meta-analysis of muon measurements in EAS with energies from PeV up to tens of EeV was presented and a remarkably consistent picture is obtained after cross-calibrating the energy scales of the different experiments. The measurements 
agree with simulations based on the models EPOS-LHC and QGSJet-II.04, within uncertainties, up to energies of a few $10^{16}\,\mathrm{eV}$, assuming the GSF flux model. At higher energies, an increasing excess with respect to model predictions is observed. The slope of this increase is between $b=0.23$ and $b=0.29$ for EPOS-LHC and between $b=0.22$ and $b=0.25$ for QGSJet-II.04, with a significance of the deviations from $b=0.0$ of around $8\sigma$ and $10\sigma$, respectively. A small change of the fits compared to the previous meta-analysis~\cite{Dembinski:2019uta,Cazon:2020zhx} is observed which arises mainly from the updates of the experimental data. Studies of other dependencies, for example the minimum energy of muons at production for each experiment, as described in Ref.~\cite{Cazon:2020zhx}, have yet been inconclusive due to the limited experimental data. It has been shown that the mass composition model is important for the mass subtraction in $\Delta z=z-z_\mathrm{mass}$ and should be derived from optical measurements with the least amount of assumptions. The GSF model satisfies this requirement at a certain level, however, the possibility to use data from $X_\mathrm{max}$ measurements directly is currently under investigation.%This level can potentially also be reached when assuming a unreasonably heavy mass composition at the highest energies, in strong tension with optical mass composition measurements based on the  maximum shower depth, $X_\mathrm{max}$.

When removing individual data from this meta-analysis, for extreme assumptions of the correlation of uncertainties, the significances for a non-zero slope can decrease to approximately $3\sigma$ for EPOS-LHC and around $5\sigma$ for QGSJet-II.04. This decrease is mainly due to the data from Yakutsk, which becomes more important when data from IceCube (or to some extent SUGAR) are removed and is in strong tension with other muon measurements. To understand these tensions, further studies of the treatment of systematic uncertainties and relative biases of the individual experiments are necessary. Moreover, additional measurements with high precision are needed, in particular at energies above \SI{e17}{eV}. Ongoing EAS detector upgrades and improved analysis methods are expected to reduce the uncertainties of the experimental results and increase the parameter space. This will improve the data accuracy and help to investigate other dependencies of the deviations between simulations and data in future extensions of this meta-analysis.

%\end{linenumbers}

\begin{spacing}{.9}

\bibliographystyle{ICRC}
\bibliography{references}

\end{spacing}

\newpage 

\section*{Full Author List:}

\input{authorlist_EAS-MSU}
\input{authorlist_IceCube}

\input{authorlist_KASCADE}
\input{authorlist_NEVOD}
\input{authorlist_Auger}
\input{authorlist_SUGAR}

\input{authorlist_TA}
\input{authorlist_Yakutsk}

\end{document}

%% file: authorlist_EAS-MSU.tex
\subsection*{EAS-MSU Collaboration:}

\scriptsize
\noindent
Yu.~A.~Fomin$^{1}$, 
N.~N.~Kalmykov$^{1}$, 
I.~S.~Karpikov$^{2}$, 
G.~V.~Kulikov$^{1}$, 
M.~Yu.~Kuznetsov$^{2,3}$, 
G.~I.~Rubtsov$^{2}$, 
V.~P.~Sulakov$^{1}$, 
S.~V.~Troitsky$^{2}$\\

\noindent
$^{1}$ D.V. Skobeltsyn Institute of Nuclear Physics, M.V. Lomonosov Moscow State University, Moscow 119991, Russia\\
$^{2}$ Institute for Nuclear Research of the Russian Academy of Sciences, Moscow, Russia\\
$^{3}$ Service de Physique Théorique, Université Libre de Bruxelles, Brussels, Belgium

%% file: authorlist_IceCube.tex
\subsection*{IceCube Collaboration:}

\scriptsize
\noindent
R. Abbasi$^{17}$,
M. Ackermann$^{59}$,
J. Adams$^{18}$,
J. A. Aguilar$^{12}$,
M. Ahlers$^{22}$,
M. Ahrens$^{50}$,
C. Alispach$^{28}$,
A. A. Alves Jr.$^{31}$,
N. M. Amin$^{42}$,
R. An$^{14}$
% ...
\\

\noindent
% ...
$^{12}$ Universit{\'e} Libre de Bruxelles, Science Faculty CP230, B-1050 Brussels, Belgium \\
% ...
$^{14}$ Department of Physics and Laboratory for Particle Physics and Cosmology, Harvard University, Cambridge, MA 02138, USA \\
% ...
$^{17}$ Department of Physics, Loyola University Chicago, Chicago, IL 60660, USA \\
$^{18}$ Dept. of Physics and Astronomy, University of Canterbury, Private Bag 4800, Christchurch, New Zealand \\
% ...
$^{22}$ Niels Bohr Institute, University of Copenhagen, DK-2100 Copenhagen, Denmark \\
% ...
$^{28}$ D{\'e}partement de physique nucl{\'e}aire et corpusculaire, Universit{\'e} de Gen{\`e}ve, CH-1211 Gen{\`e}ve, Switzerland \\
% ...
$^{31}$ Karlsruhe Institute of Technology, Institute for Astroparticle Physics, D-76021 Karlsruhe, Germany  \\
% ...
$^{42}$ Bartol Research Institute and Dept. of Physics and Astronomy, University of Delaware, Newark, DE 19716, USA \\
% ...
$^{50}$ Oskar Klein Centre and Dept. of Physics, Stockholm University, SE-10691 Stockholm, Sweden \\
% ...
$^{59}$ DESY, D-15738 Zeuthen, Germany \\
% ...

\noindent{\fontsize{10}{10}\selectfont\bf Acknowledgements:}

\vspace{0.5em}

\noindent
USA {\textendash} U.S. National Science Foundation-Office of Polar Programs,
U.S. National Science Foundation-Physics Division,
U.S. National Science Foundation-EPSCoR,
Wisconsin Alumni Research Foundation,
Center for High Throughput Computing (CHTC) at the University of Wisconsin{\textendash}Madison,
Open Science Grid (OSG),
Extreme Science and Engineering Discovery Environment (XSEDE),
Frontera computing project at the Texas Advanced Computing Center,
U.S. Department of Energy-National Energy Research Scientific Computing Center,
Particle astrophysics research computing center at the University of Maryland,
Institute for Cyber-Enabled Research at Michigan State University,
and Astroparticle physics computational facility at Marquette University;
Belgium {\textendash} Funds for Scientific Research (FRS-FNRS and FWO),
FWO Odysseus and Big Science programmes,
and Belgian Federal Science Policy Office (Belspo);
Germany {\textendash} Bundesministerium f{\"u}r Bildung und Forschung (BMBF),
Deutsche Forschungsgemeinschaft (DFG),
Helmholtz Alliance for Astroparticle Physics (HAP),
Initiative and Networking Fund of the Helmholtz Association,
Deutsches Elektronen Synchrotron (DESY),
and High Performance Computing cluster of the RWTH Aachen;
Sweden {\textendash} Swedish Research Council,
Swedish Polar Research Secretariat,
Swedish National Infrastructure for Computing (SNIC),
and Knut and Alice Wallenberg Foundation;
Australia {\textendash} Australian Research Council;
Canada {\textendash} Natural Sciences and Engineering Research Council of Canada,
Calcul Qu{\'e}bec, Compute Ontario, Canada Foundation for Innovation, WestGrid, and Compute Canada;
Denmark {\textendash} Villum Fonden and Carlsberg Foundation;
New Zealand {\textendash} Marsden Fund;
Japan {\textendash} Japan Society for Promotion of Science (JSPS)
and Institute for Global Prominent Research (IGPR) of Chiba University;
Korea {\textendash} National Research Foundation of Korea (NRF);
Switzerland {\textendash} Swiss National Science Foundation (SNSF);
United Kingdom {\textendash} Department of Physics, University of Oxford.\\

%% file: authorlist_KASCADE.tex
\subsection*{KASCADE-Grande Collaboration:}

\scriptsize
\noindent
W. D. Apel$^{1}$, 
J. C. Arteaga-Vel\'azquez$^{2}$, 
K. Bekk$^{1}$, 
M. Bertaina$^{3}$,
J. Bl\"umer$^{1,4}$, 
H. Bozdog$^{1}$, 
E. Cantoni$^{3,5}$, 
A. Chiavassa$^{3}$,
F. Cossavella$^{4}$, 
K. Daumiller$^{1}$, 
V. de Souza$^{7}$, 
F. Di Pierro$^{3}$, 
P. Doll$^{1}$,
R. Engel$^{1,4}$, 
D. Fuhrmann$^{8}$, 
A. Gherghel-Lascu$^{5}$, 
H. J. Gils$^{1}$, 
R. Glasstetter$^{8}$,
C. Grupen$^{9}$, 
A. Haungs$^{1}$, 
D. Heck$^{1}$, 
J. R. H\"orandel$^{10}$, 
T. Huege$^{1}$,
K.-H. Kampert$^{8}$, 
D. Kang$^{1}$,
H. O. Klages$^{1}$, 
K. Link$^{1}$, 
P. {\L}uczak$^{11}$, 
H. J. Mathes$^{1}$,
H. J. Mayer$^{1}$, 
J. Milke$^{1}$, 
C. Morello$^{6}$, 
J. Oehlschl\"ager$^{1}$,
S. Ostapchenko$^{12}$, 
T. Pierog$^{1}$, 
H. Rebel$^{1}$, 
D. Rivera-Rangel$^{2}$, 
M. Roth$^{1}$,
H. Schieler$^{1}$, 
S. Schoo$^{1}$, 
F. G. Schr\"oder$^{1}$, 
O. Sima$^{13}$, 
G. Toma$^{5}$,
G. C. Trinchero$^{6}$, 
H. Ulrich$^{1}$, 
A. Weindl$^{1}$, 
J. Wochele$^{1}$, 
J. Zabierowski$^{11}$\\

\noindent
$^{1}$ Karlsruhe Institute of Technology, Institute for Astroparticle Physics, Karlsruhe, Germany\\
$^{2}$ Universidad Michoacana, Inst.~F\'{\i}sica y Matem\'aticas, Morelia, Mexico\\
$^{3}$ Dipartimento di Fisica, Universit\`a degli Studi di Torino, Italy\\
$^{4}$ Institut f\"ur Experimentelle Teilchenphysik KIT - Karlsruhe Institute of Technology, Germany\\
$^{5}$ Horia Hulubei National Institute of Physics and Nuclear Engineering, Bucharest, Romania\\
$^{6}$ Osservatorio Astrofisico di Torino, INAF Torino, Italy\\
$^{7}$ Universidade S$\tilde{a}$o Paulo, Instituto de F\'{\i}sica de S\~ao Carlos, Brasil\\
$^{8}$ Fachbereich Physik, Universit\"at Wuppertal, Germany\\
$^{9}$ Department of Physics, Siegen University, Germany\\
$^{10}$ Dept. of Astrophysics, Radboud University Nijmegen, The Netherlands\\
$^{11}$ National Centre for Nuclear Research, Department of Astrophysics, Lodz, Poland\\
$^{12}$ Frankfurt Institute for Advanced Studies (FIAS), Frankfurt am Main, Germany\\
$^{13}$ Department of Physics, University of Bucharest, Bucharest, Romania\\

%% file: authorlist_NEVOD.tex
\subsection*{NEVDOD-DECOR Collaboration:}

\scriptsize
\noindent 
N. S. Barbashina$^{1}$, 
A. G. Bogdanov$^{1}$, 
S. S. Khokhlov$^{1}$, 
V. V. Kindin$^{1}$,
R. P. Kokoulin$^{1}$,
K. G. Kompaniets$^{1}$,
G. Mannocchi$^{2}$,
A. A. Petrukhin$^{1}$,
V. V. Shutenko$^{1}$,
G. Trinchero$^{2}$,
I. I. Yashin$^{1}$,
E. A. Yurina$^{1}$,
E. A. Zadeba$^{1}$\\

\noindent
$^{1}$ National Research Nuclear University MEPhI (Moscow Engineering Physics Institute), Moscow, Russia\\
$^{2}$ Osservatorio Astrofisico di Torino – INAF, Torino, Italy\\

\noindent{\fontsize{10}{10}\selectfont\bf Acknowledgements:}

\vspace{0.5em}

\noindent
The work of the NEVOD-DECOR group is supported by the Ministry of Science and Higher Education of the Russian Federation, project “Fundamental problems of cosmic rays and dark matter”, No. 0723-2020-0040.

%% file: authorlist_Auger.tex
\subsection*{Pierre Auger Collaboration:}

\scriptsize
\noindent
P.~Abreu$^{72}$,
M.~Aglietta$^{54,52}$,
J.M.~Albury$^{13}$,
I.~Allekotte$^{1}$,
A.~Almela$^{8,12}$,
J.~Alvarez-Mu\~niz$^{79}$,
R.~Alves Batista$^{80}$,
G.A.~Anastasi$^{63,52}$,
L.~Anchordoqui$^{87}$,
B.~Andrada$^{8}$
% ...
\\

\noindent
$^{1}$ Centro At\'omico Bariloche and Instituto Balseiro (CNEA-UNCuyo-CONICET), San Carlos de Bariloche, Argentina\\
% ...
$^{8}$ Instituto de Tecnolog\'\i{}as en Detecci\'on y Astropart\'\i{}culas (CNEA, CONICET, UNSAM), Buenos Aires, Argentina\\
% ...
$^{12}$ Universidad Tecnol\'ogica Nacional -- Facultad Regional Buenos Aires, Buenos Aires, Argentina\\
$^{13}$ University of Adelaide, Adelaide, S.A., Australia\\
% ...
$^{52}$ INFN, Sezione di Torino, Torino, Italy\\
% ...
$^{54}$ Osservatorio Astrofisico di Torino (INAF), Torino, Italy\\
% ...
$^{63}$ Universit\`a Torino, Dipartimento di Fisica, Torino, Italy\\
% ...
$^{72}$ Laborat\'orio de Instrumenta\c{c}\~ao e F\'\i{}sica Experimental de Part\'\i{}culas -- LIP and Instituto Superior T\'ecnico -- IST, Universidade de Lisboa -- UL, Lisboa, Portugal\\
% ...
$^{79}$ Instituto Galego de F\'\i{}sica de Altas Enerx\'\i{}as (IGFAE), Universidade de Santiago de Compostela, Santiago de Compostela, Spain\\
$^{80}$ IMAPP, Radboud University Nijmegen, Nijmegen, The Netherlands\\
% ...
$^{87}$ Department of Physics and Astronomy, Lehman College, City University of New York, Bronx, NY, USA\\
% ...
$^{c}$ School of Physics and Astronomy, University of Leeds, Leeds, United Kingdom\\
% ...

\noindent{\fontsize{10}{10}\selectfont\bf Acknowledgements:}

\vspace{0.5em}

\noindent

The successful installation, commissioning, and operation of the Pierre
Auger Observatory would not have been possible without the strong
commitment and effort from the technical and administrative staff in
Malarg\"ue. We are very grateful to the following agencies and
organizations for financial support:

Argentina -- Comisi\'on Nacional de Energ\'\i{}a At\'omica; Agencia Nacional de
Promoci\'on Cient\'\i{}fica y Tecnol\'ogica (ANPCyT); Consejo Nacional de
Investigaciones Cient\'\i{}ficas y T\'ecnicas (CONICET); Gobierno de la
Provincia de Mendoza; Municipalidad de Malarg\"ue; NDM Holdings and Valle
Las Le\~nas; in gratitude for their continuing cooperation over land
access; Australia -- the Australian Research Council; Belgium -- Fonds
de la Recherche Scientifique (FNRS); Research Foundation Flanders (FWO);
Brazil -- Conselho Nacional de Desenvolvimento Cient\'\i{}fico e Tecnol\'ogico
(CNPq); Financiadora de Estudos e Projetos (FINEP); Funda\c{c}\~ao de Amparo \`a
Pesquisa do Estado de Rio de Janeiro (FAPERJ); S\~ao Paulo Research
Foundation (FAPESP) Grants No.~2019/10151-2, No.~2010/07359-6 and
No.~1999/05404-3; Minist\'erio da Ci\^encia, Tecnologia, Inova\c{c}\~oes e
Comunica\c{c}\~oes (MCTIC); Czech Republic -- Grant No.~MSMT CR LTT18004,
LM2015038, LM2018102, CZ.02.1.01/0.0/0.0/16{\textunderscore}013/0001402,
CZ.02.1.01/0.0/0.0/18{\textunderscore}046/0016010 and
CZ.02.1.01/0.0/0\\.0/17{\textunderscore}049/0008422; France -- Centre de Calcul
IN2P3/CNRS; Centre National de la Recherche Scientifique (CNRS); Conseil
R\'egional Ile-de-France; D\'epartement Physique Nucl\'eaire et Corpusculaire
(PNC-IN2P3/CNRS); D\'epartement Sciences de l'Univers (SDU-INSU/CNRS);
Institut Lagrange de Paris (ILP) Grant No.~LABEX ANR-10-LABX-63 within
the Investissements d'Avenir Programme Grant No.~ANR-11-IDEX-0004-02;
Germany -- Bundesministerium f\"ur Bildung und Forschung (BMBF); Deutsche
Forschungsgemeinschaft (DFG); Finanzministerium Baden-W\"urttemberg;
Helmholtz Alliance for Astroparticle Physics (HAP);
Helmholtz-Gemeinschaft Deutscher Forschungszentren (HGF); Ministerium
f\"ur Innovation, Wissenschaft und Forschung des Landes
Nordrhein-Westfalen; Ministerium f\"ur Wissenschaft, Forschung und Kunst
des Landes Baden-W\"urttemberg; Italy -- Istituto Nazionale di Fisica
Nucleare (INFN); Istituto Nazionale di Astrofisica (INAF); Ministero
dell'Istruzione, dell'Universit\'a e della Ricerca (MIUR); CETEMPS Center
of Excellence; Ministero degli Affari Esteri (MAE); M\'exico -- Consejo
Nacional de Ciencia y Tecnolog\'\i{}a (CONACYT) No.~167733; Universidad
Nacional Aut\'onoma de M\'exico (UNAM); PAPIIT DGAPA-UNAM; The Netherlands
-- Ministry of Education, Culture and Science; Netherlands Organisation
for Scientific Research (NWO); Dutch national e-infrastructure with the
support of SURF Cooperative; Poland -- Ministry of Education and
Science, grant No.~DIR/WK/2018/11; National Science Centre, Grants
No.~2016/22/M/ST9/00198, 2016/23/B/ST9/01635, and 2020/39/B/ST9/01398;
Portugal -- Portuguese national funds and FEDER funds within Programa
Operacional Factores de Competitividade through Funda\c{c}\~ao para a Ci\^encia
e a Tecnologia (COMPETE); Romania -- Ministry of Research, Innovation
and Digitization, CNCS/CCCDI -- UEFISCDI, projects PN19150201/16N/2019,
PN1906010, TE128 and PED289, within PNCDI III; Slovenia -- Slovenian
Research Agency, grants P1-0031, P1-0385, I0-0033, N1-0111; Spain --
Ministerio de Econom\'\i{}a, Industria y Competitividad (FPA2017-85114-P and
PID2019-104676GB-C32), Xunta de Galicia (ED431C 2017/07), Junta de
Andaluc\'\i{}a (SOMM17/6104/UGR, P18-FR-4314) Feder Funds, RENATA Red
Nacional Tem\'atica de Astropart\'\i{}culas (FPA2015-68783-REDT) and Mar\'\i{}a de
Maeztu Unit of Excellence (MDM-2016-0692); USA -- Department of Energy,
Contracts No.~DE-AC02-07CH11359, No.~DE-FR02-04ER41300,
No.~DE-FG02-99ER41107 and No.~DE-SC0011689; National Science Foundation,
Grant No.~0450696; The Grainger Foundation; Marie Curie-IRSES/EPLANET;
European Particle Physics Latin American Network; and UNESCO.

%% file: authorlist_SUGAR.tex
\subsection*{SUGAR Collaboration:}

\scriptsize
\noindent
N.~N.~Kalmykov$^{1}$, 
I.~S.~Karpikov$^{2}$, 
G.~I.~Rubtsov$^{2}$, 
S.~V.~Troitsky$^{2}$,
J. Ulrichs$^{3}$\\

\noindent
$^{1}$ D.V. Skobeltsyn Institute of Nuclear Physics, M.V. Lomonosov Moscow State University, Moscow 119991, Russia\\
$^{2}$ Institute for Nuclear Research of the Russian Academy of Sciences, Moscow, Russia\\
$^{3}$ jusyd19@gmail.com

%% file: authorlist_TA.tex
\subsection*{Telescope Array Collaboration:}

\scriptsize
\noindent
R.~U.~Abbasi$^{1}$,
M.~Abe$^{2}$,
T.~Abu-Zayyad$^{1,3}$,
M.~Allen$^{3}$,
Y.~Arai$^{4}$,
E.~Barcikowski$^{3}$,
J.~W.~Belz$^{3}$,
D.~R.~Bergman$^{3}$,
S.~A.~Blake$^{3}$,
I.~Buckland$^{3}$
% ...
\\

\noindent
$^{1}$ Department of Physics, Loyola University Chicago, Chicago, Illinois, USA \\
$^{2}$ The Graduate School of Science and Engineering, Saitama University, Saitama, Saitama, Japan \\
$^{3}$ High Energy Astrophysics Institute and Department of Physics and Astronomy, University of Utah, Salt Lake City, Utah, USA \\
$^{4}$ Graduate School of Science, Osaka City University, Osaka, Osaka, Japan \\
% ...

\noindent{\fontsize{10}{10}\selectfont\bf Acknowledgements:}

\vspace{0.5em}

\noindent The Telescope Array experiment is supported by the Japan Society for
the Promotion of Science(JSPS) through
Grants-in-Aid
for Priority Area
%"Highest Energy Cosmic Rays"
431,
for Specially Promoted Research
%``Extreme Phenomena in the Universe Explored by Highest Energy Cosmic Rays''
%Grant Number
JP21000002,
%Grant-in-Aid
for Scientific  Research (S)
%"Quest for the unified picture of the explosion mechanism of supernovae and the central engine of gamma-ray bursts"
%Grant Number
JP19104006,
%Grant-in-Aid
for Specially Promoted Research
%"Extended Telescope Array Experiment - Nearby Extreme Universe Elucidated by Highest-energy Cosmic Rays"
%Grant Number
JP15H05693,
%Grant-in-Aid
for Scientific  Research (S)
%"Study of the ultra high energy cosmic ray source evolution by detailed measurement of cosmic rays in the wide energy range"
%Grant Number
JP15H05741 and JP19H05607,
 for Science Research (A) JP18H03705,
%Grant-in-Aid
for Young Scientists (A)
%"hoge hoge"
%Grant Number
JPH26707011,
and for Fostering Joint International Research (B)
%"Search for Ultra-High Energy Cosmic Ray origin using the extended Telescope Array experiment"
%Grant Number
JP19KK0074,
by the joint research program of the Institute for Cosmic Ray Research (ICRR), The University of Tokyo;
by the Pioneering Program of RIKEN for the Evolution of Matter in the Universe (r-EMU);
by the U.S. National Science
Foundation awards PHY-1404495, PHY-1404502, PHY-1607727, PHY-1712517, PHY-1806797 and PHY-2012934;
by the National Research Foundation of Korea
% \linebreak
(2017K1A4A3015188, 2020R1A2C1008230, \& 2020R1A2C2102800) ;
%\linebreak
by the Ministry of Science and Higher Education of the Russian Federation under the contract 075-15-2020-778, RFBR grant 20-02-00625a (INR), IISN project No. 4.4501.18, and Belgian Science Policy under IUAP VII/37 (ULB). This work was partially supported by the grants ofThe joint research program of the Institute for Space-Earth Environmental Research, Nagoya University and Inter-University Research Program of the Institute for Cosmic Ray Research of University of Tokyo. The foundations of Dr. Ezekiel R. and Edna Wattis Dumke, Willard L. Eccles, and George S. and Dolores Dor\'e Eccles all helped with generous donations. The State of Utah supported the project through its Economic Development Board, and the University of Utah through the Office of the Vice President for Research. The experimental site became available through the cooperation of the Utah School and Institutional Trust Lands Administration (SITLA), U.S. Bureau of Land Management (BLM), and the U.S. Air Force. We appreciate the assistance of the State of Utah and Fillmore offices of the BLM in crafting the Plan of Development for the site.  Patrick A.~Shea assisted the collaboration with valuable advice and supported the collaboration’s efforts. The people and the officials of Millard County, Utah have been a source of steadfast and warm support for our work which we greatly appreciate. We are indebted to the Millard County Road Department for their efforts to maintain and clear the roads which get us to our sites. We gratefully acknowledge the contribution from the technical staffs of our home institutions. An allocation of computer time from the Center for High Performance Computing at the University of Utah is gratefully acknowledged.

%% file: authorlist_Yakutsk.tex
\subsection*{Yakutsk EAS Array Collaboration:}

\scriptsize
\noindent
E.~A.~Atlasov$^{1}$,
N.~G.~Bolotnikov$^{1}$,
N.~A.~Dyachkovskiy$^{1}$,
N.~S.~Gerasimova$^{1}$,
A.~V.~Glushkov$^{1}$,
A.~A.~Ivanov$^{1}$,
O.~N.~Ivanov$^{1}$,
I.~A.~Kellarev$^{1}$,
S.~P.~Knurenko$^{1}$,
A.~D.~Krasilnikov$^{1}$,
I.~V.~Ksenofontov$^{1}$,
L.~T.~Ksenofontov$^{1}$,
K.~G.~Lebedev$^{1}$,
S.~V.~Matarkin$^{1}$,
V.~P.~Mokhnachevskaya$^{1}$, 
N.~I.~Neustroev$^{1}$,
I.~S.~Petrov$^{1}$,
A.~S.~Proshutinsky$^{1}$, 
A.~V.~Sabourov$^{1}$,
I.~Ye.~Sleptsov$^{1}$,
G.~G.~Struchkov$^{1}$,
L.~V.~Timofeev$^{1}$,
B.~B.~Yakovlev$^{1}$\\

\noindent
$^{1}$ Yu. G. Shafer Institute of Cosmophysical Research and Aeronomy SB RAS 677980 Yakutsk, Russia\\

\noindent{\fontsize{10}{10}\selectfont\bf Acknowledgements:}

\vspace{0.5em}

\noindent
The Yakutsk EAS Array is supported by the Ministry of Science and Higher Education of the Russian Federation under the contract AAAA-A21-121011990011-8.

%% file: skeleton.bbl
\providecommand{\href}[2]{#2}\begingroup\raggedright\begin{thebibliography}{10}

\bibitem{Kampert:2012mx}
K.-H. Kampert and M.~Unger
  \href{http://dx.doi.org/10.1016/j.astropartphys.2012.02.004}{{\em Astropart.
  Phys.} {\bfseries 35} (2012) 660--678}.

\bibitem{BeckerTjus:2020xzg}
J.~Becker~Tjus and L.~Merten
  \href{http://dx.doi.org/10.1016/j.physrep.2020.05.002}{{\em Phys. Rept.}
  {\bfseries 872} (2020) 1--98}.

\bibitem{AbuZayyad:1999xa}
{\bfseries HiRes-MIA} Collaboration, T.~Abu-Zayyad {\em et~al.}
  \href{http://dx.doi.org/10.1103/PhysRevLett.84.4276}{{\em Phys. Rev. Lett.}
  {\bfseries 84} (2000) 4276--4279}.

\bibitem{Bogdanov:2010zz}
{\bfseries NEVOD-DECOR} Collaboration, A.~G. Bogdanov {\em et~al.}
  \href{http://dx.doi.org/10.1134/S1063778810110074}{{\em Phys. Atom. Nucl.}
  {\bfseries 73} (2010) }.

\bibitem{Bogdanov:2018sfw}
{\bfseries NEVOD-DECOR} Collaboration, A.~G. Bogdanov {\em et~al.}
  \href{http://dx.doi.org/10.1016/j.astropartphys.2018.01.003}{{\em Astropart.
  Phys.} {\bfseries 98} (2018) 13--20}.

\bibitem{Aab:2014pza}
{\bfseries Pierre Auger} Collaboration, A.~Aab {\em et~al.}
  \href{http://dx.doi.org/10.1103/PhysRevD.91.032003}{{\em Phys. Rev. D}
  {\bfseries 91} no.~3, (2015) 032003}.

\bibitem{Aab:2016hkv}
{\bfseries Pierre Auger} Collaboration, A.~Aab {\em et~al.}
  \href{http://dx.doi.org/10.1103/PhysRevLett.117.192001}{{\em Phys. Rev.
  Lett.} {\bfseries 117} no.~19, (2016) 192001}.

\bibitem{Abbasi:2018fkz}
{\bfseries Telescope Array} Collaboration, R.~U. Abbasi {\em et~al.}
  \href{http://dx.doi.org/10.1103/PhysRevD.98.022002}{{\em Phys. Rev. D}
  {\bfseries 98} no.~2, (2018) 022002}.

\bibitem{Bellido:2018toz}
{\bfseries SUGAR} Collaboration, J.~A. Bellido {\em et~al.}
  \href{http://dx.doi.org/10.1103/PhysRevD.98.023014}{{\em Phys. Rev. D}
  {\bfseries 98} no.~2, (2018) 023014}.

\bibitem{Fomin:2016kul}
{\bfseries EAS-MSU} Collaboration, Y.~A. Fomin {\em et~al.}
  \href{http://dx.doi.org/10.1016/j.astropartphys.2017.04.001}{{\em Astropart.
  Phys.} {\bfseries 92} (2017) 1--6}.

\bibitem{Glushkov}
{\bfseries Yakutsk} Collaboration, A.~Glushkov, M.~Pravdin, and A.~Sabourov
  {\em Priv. Comm.} (2018) .

\bibitem{Apel:2017thr}
{\bfseries KASCADE-Grande} Collaboration, W.~D. Apel {\em et~al.}
  \href{http://dx.doi.org/10.1016/j.astropartphys.2017.07.001}{{\em Astropart.
  Phys.} {\bfseries 95} (2017) 25--43}.

\bibitem{Dembinski:2019uta}
{\bfseries EAS-MSU, IceCube, KASCADE Grande, NEVOD-DECOR, Pierre Auger, SUGAR,
  Telescope Array, Yakutsk EAS Array} Collaboration, H.~P. Dembinski {\em
  et~al.} \href{http://dx.doi.org/10.1051/epjconf/201921002004}{{\em EPJ Web
  Conf.} {\bfseries 210} (2019) 02004}.

\bibitem{Cazon:2020zhx}
{\bfseries EAS-MSU, IceCube, KASCADE Grande, NEVOD-DECOR, Pierre Auger, SUGAR,
  Telescope Array, Yakutsk EAS Array} Collaboration, L.~Cazon
  \href{http://dx.doi.org/10.22323/1.358.0214}{{\em PoS} {\bfseries ICRC2019}
  (2020) 214}.

\bibitem{Aab:2021zfr}
{\bfseries Pierre Auger} Collaboration, A.~Aab {\em et~al.}
  \href{http://dx.doi.org/10.1103/PhysRevLett.126.152002}{{\em Phys. Rev.
  Lett.} {\bfseries 126} no.~15, (2021) 152002}.

\bibitem{Aab:2020frk}
{\bfseries Pierre Auger} Collaboration, A.~Aab {\em et~al.}
  \href{http://dx.doi.org/10.1140/epjc/s10052-020-8055-y}{{\em Eur. Phys. J. C}
  {\bfseries 80} no.~8, (2020) 751}.

\bibitem{Soldin:2021}
{\bfseries IceCube} Collaboration, D.~Soldin
  \href{http://pos.sissa.it/395/342/}{{\em PoS} {\bfseries ICRC2021} (2021)
  342}.

\bibitem{Gesualdi:ICRC2021}
F.~Gesualdi {\em et~al.} \href{http://doi.org/10.22323/1.395.0473}{{\em PoS}
  {\bfseries ICRC2021} (2021) 473}.

\bibitem{Albrecht:2021yla}
J.~Albrecht {\em et~al.} \href{https://arxiv.org/abs/2105.06148}{{\em submitted
  to Astrophysics and Space Science} (2021) }.

\bibitem{Pierog:2013ria}
T.~Pierog {\em et~al.} \href{http://dx.doi.org/10.1103/PhysRevC.92.034906}{{\em
  Phys. Rev. C} {\bfseries 92} no.~3, (2015) 034906}.

\bibitem{Ostapchenko:2013pia}
S.~Ostapchenko \href{http://dx.doi.org/10.1051/epjconf/20125202001}{{\em EPJ
  Web Conf.} {\bfseries 52} (2013) 02001}.

\bibitem{Riehn:2017mfm}
F.~Riehn {\em et~al.} \href{http://dx.doi.org/10.22323/1.301.0301}{{\em PoS}
  {\bfseries ICRC2017} (2018) 301}.

\bibitem{Engel:2019dsg}
F.~Riehn {\em et~al.} \href{http://dx.doi.org/10.1103/PhysRevD.102.063002}{{\em
  Phys. Rev. D} {\bfseries 102} no.~6, (2020) 063002}.

\bibitem{Ostapchenko:1993}
N.~Kalmykov and S.~Ostapchenko {\em Phys. Atom. Nucl.} {\bfseries 56} no.~346,
  (1993) .

\bibitem{Ahn:2009wx}
E.-J. Ahn {\em et~al.} \href{http://dx.doi.org/10.1103/PhysRevD.80.094003}{{\em
  Phys. Rev. D} {\bfseries 80} (2009) 094003}.

\bibitem{Matthews:2005sd}
J.~Matthews \href{http://dx.doi.org/10.1016/j.astropartphys.2004.09.003}{{\em
  Astropart. Phys.} {\bfseries 22} (2005) 387--397}.

\bibitem{Dembinski:2015xtn}
H.~P. Dembinski {\em et~al.} \href{http://dx.doi.org/10.22323/1.301.0533}{{\em
  PoS} {\bfseries ICRC2017} (2018) 533}.

\bibitem{Deligny:2020gzq}
{\bfseries Pierre Auger, Telescope Array} Collaboration, O.~Deligny
  \href{http://dx.doi.org/10.22323/1.358.0234}{{\em PoS} {\bfseries ICRC2019}
  (2020) 234}.

\bibitem{Gaisser:2013bla}
T.~K. Gaisser, T.~Stanev, and S.~Tilav
  \href{http://dx.doi.org/10.1007/s11467-013-0319-7}{{\em Front.
  Phys.(Beijing)} {\bfseries 8} (2013) 748--758}.

\bibitem{Gaisser:2011cc}
T.~K. Gaisser \href{http://dx.doi.org/10.1016/j.astropartphys.2012.02.010}{{\em
  Astropart. Phys.} {\bfseries 35} (2012) 801--806}.

\end{thebibliography}\endgroup
